\documentclass[aps,prl,twocolumn,superscriptaddress,amsmath,amssymb]{revtex4-2}

\usepackage{graphicx,hyperref}
\usepackage{epstopdf}
\usepackage{dcolumn}
\usepackage{bm}
\usepackage{color}
\usepackage{amsmath}
\usepackage{ulem}
\usepackage{textcomp}

\begin{document}

\title{Direct visualization of electric current induced dipoles of atomic impurities}%

\affiliation{State Key Laboratory of Low-Dimensional Quantum Physics, Department of physics, Tsinghua University, Beijing 100084, China}
\affiliation{Beijing Academy of Quantum Information Sciences, Beijing 100193, China}
\affiliation{Department of Physics, Southern University of Science and Technology, Shenzhen 518055, China}
\affiliation{Frontier Science Center for Quantum Information, Beijing 100084, China}

\author{Yaowu Liu}
\thanks{These two authors contributed equally}
\affiliation{State Key Laboratory of Low-Dimensional Quantum Physics, Department of physics, Tsinghua University, Beijing 100084, China}%

\author{Zichun Zhang}
\thanks{These two authors contributed equally}
\affiliation{State Key Laboratory of Low-Dimensional Quantum Physics, Department of physics, Tsinghua University, Beijing 100084, China}%

\author{Sidan Chen}
\affiliation{State Key Laboratory of Low-Dimensional Quantum Physics, Department of physics, Tsinghua University, Beijing 100084, China}%

\author{Shengnan Xu}
\affiliation{State Key Laboratory of Low-Dimensional Quantum Physics, Department of physics, Tsinghua University, Beijing 100084, China}%

\author{Lichen Ji}
\affiliation{State Key Laboratory of Low-Dimensional Quantum Physics, Department of physics, Tsinghua University, Beijing 100084, China}%

\author{Wei Chen}
\affiliation{State Key Laboratory of Low-Dimensional Quantum Physics, Department of physics, Tsinghua University, Beijing 100084, China}%

\author{Xinyu Zhou}
\affiliation{State Key Laboratory of Low-Dimensional Quantum Physics, Department of physics, Tsinghua University, Beijing 100084, China}%

\author{Jiaxin Luo}
\affiliation{State Key Laboratory of Low-Dimensional Quantum Physics, Department of physics, Tsinghua University, Beijing 100084, China}%

\author{Xiaopen Hu}
\affiliation{State Key Laboratory of Low-Dimensional Quantum Physics, Department of physics, Tsinghua University, Beijing 100084, China}%

\author{Wenhui Duan}
\affiliation{State Key Laboratory of Low-Dimensional Quantum Physics, Department of physics, Tsinghua University, Beijing 100084, China}%

\author{Xi Chen}
\affiliation{State Key Laboratory of Low-Dimensional Quantum Physics, Department of physics, Tsinghua University, Beijing 100084, China}%

\author{Qi-Kun Xue}
\affiliation{State Key Laboratory of Low-Dimensional Quantum Physics, Department of physics, Tsinghua University, Beijing 100084, China}%
\affiliation{Beijing Academy of Quantum Information Sciences, Beijing 100193, China}
\affiliation{Department of Physics, Southern University of Science and Technology, Shenzhen 518055, China}
\affiliation{Frontier Science Center for Quantum Information, Beijing 100084, China}

\author{Shuai-Hua Ji}
\email{shji@mail.tsinghua.edu.cn}
\affiliation{State Key Laboratory of Low-Dimensional Quantum Physics, Department of physics, Tsinghua University, Beijing 100084, China}
\affiliation{Frontier Science Center for Quantum Information, Beijing 100084, China}

\date{\today}%

\begin{abstract}
Learning the electron scattering around atomic impurities is a fundamental step to fully understand the basic electronic transport properties of realistic conducting materials.
Although many efforts have been made in this field for several decades~\cite{muralt1986scanning,muralt1987scanning,chu1989scanning,kirtley1988direct,bannani2008local,krebs2023imaging,homoth2009electronic}, atomic scale transport around single point-like impurities has yet been achieved. Here, we report the direct visualization of the electric current induced dipoles around single atomic impurities in epitaxial bilayer graphene by multi-probe low temperature scanning tunneling potentiometry as the local current density is raised up to around 25 A/m, which is considerably higher than that in previous studies~\cite{ji2012atomic,willke2015spatial}.
We find the directions of these dipoles which are parallel or anti-parallel to local current are determined by the charge polarity of the impurities, revealing the direct evidence for the existence of the carrier density modulation effect proposed by Landauer in 1976~\cite{landauer1974driving,landauer1976spatial}.
Furthermore, by $in$ $situ$ tuning local current direction with contact probes, these dipoles are redirected correspondingly.
Our work paves the way to explore the electronic quantum transport phenomena at single atomic impurity level and the potential future electronics toward or beyond the end of Moore's Law.
\end{abstract}

\maketitle

\section{Introduction}\label{sec1}
Quantum transport properties of electron in real materials with unavoidable defects or impurities are essential for modern electronics. However, the semiclassical theory based on Boltzmann equation, where only average current and potential distribution can be achieved, is not adequate to describe microscopic process of electronic quantum transport phenomenon around atomic defect~\cite{bevan2014first,morr2017scanning}. The microscopic transport theory proposed by Landauer \cite{landauer1957spatial,landauer1974driving,landauer1976spatial} captures the potential fluctuation, which is the long sought-after current induced dipole, at atomic scale near individual impurities.

\begin{figure*}
   \begin{center}
   \includegraphics[width=0.6\textwidth]{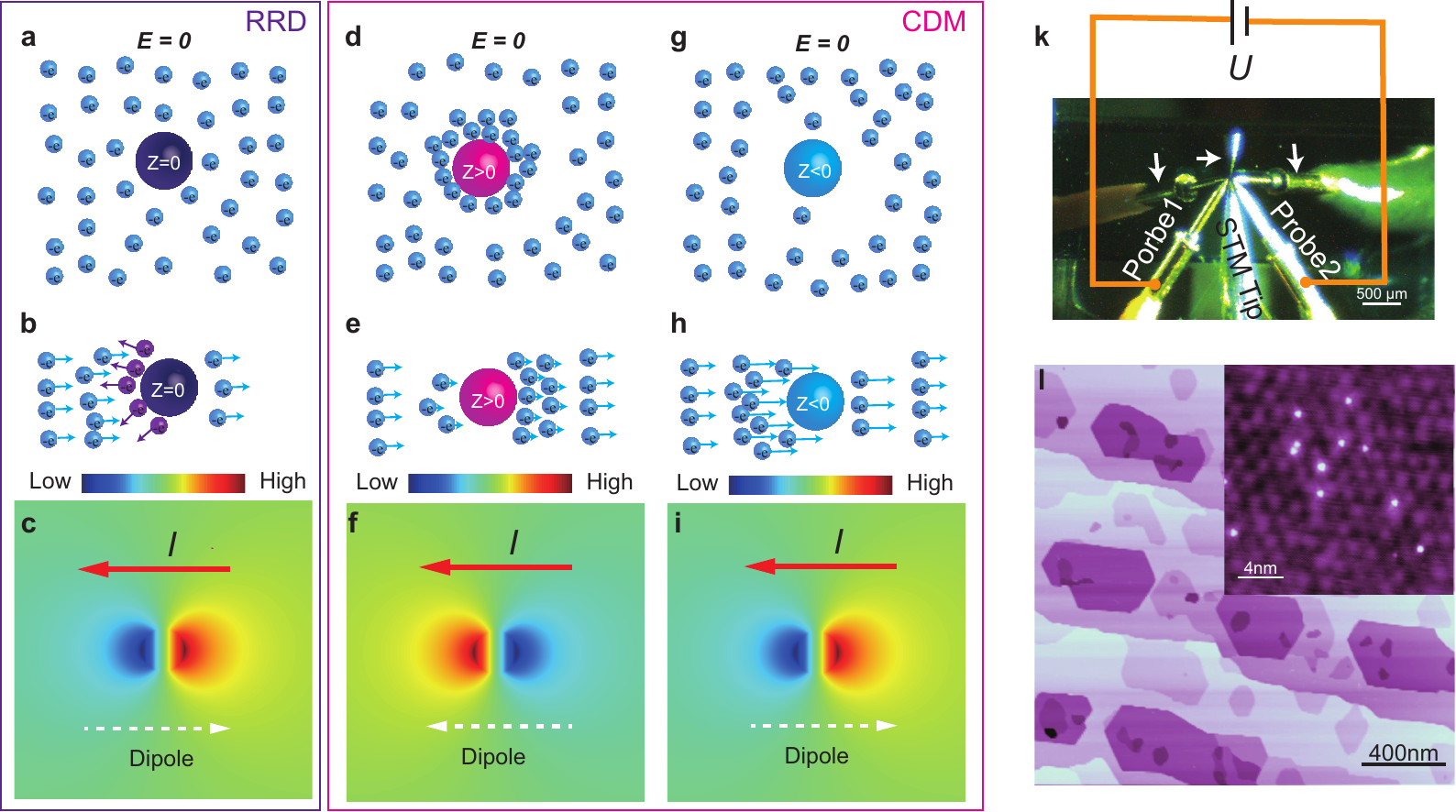}
   \end{center}
    \caption{ Current induced dipoles and STP configuration. a-c, Schematics for RRD. Uniform electron distribution around chargeless impurity (a). Electrons move under electric field and scattered by a chargeless impurity; the blue spheres represent forward moving electrons and the purple spheres represent scattered electrons (b). Potential distribution around chargeless impurities, the direction of the dipole (the dashed white arrow) is anti-aligned with local current (c). d-i, Schematics for CDM effects of both positively and negatively charged impurities. Electron distribution disturbed by charged impurities (d,g). Electrons velocity are modified due to continuity of current near the impurities, with the length of the arrow standing for the magnitude of velocity (e,h). Potential distribution around charged impurities (f,i). For positively charged impurity, the direction of the dipole is parallel with local current (f) while for negatively charged impurity the direction of the dipole is anti-parallel with local current (i). k, Optical picture of tip arrangement for STP measurement. The white arrows mark the reflection images of the probes. l, Typical topography of nitrogen doped bilayer graphene, 2 $\rm \times$ 2 \textmu m, $V_{\rm b}$ = 1 V, $I_{\rm t}$ = 10 pA. Inset: topography of bilayer graphene with some nitrogen substitutes shown as bright protrusions, 20 $\rm \times$ 20 nm, $V_{\rm b}$ = 100 mV, $I_{\rm t}$ = 200 pA.}\label{Fig1}
\end{figure*}

Two types of current induced dipoles have been theoretically predicted: one is the residual resistive dipole (RRD), and the other is the carrier density modulation (CDM) dipole. The RRD is formed when the carrier concentration on one side of back scattering center is increased and the other side is reduced as shown in Fig.~\ref{Fig1}(a-c). RRD has been adapted to explain the ultimate resistance of metallic materials at low temperature (Matthiessen's Rule)~\cite{landauer1957spatial}. In two dimensional non-interacting electron systems the RRD moment is given exactly by~\cite{zwerger1991exact}:
\begin{equation}\label{EQ1}
\vec{p}=\frac{\hbar k_{\rm F} \vec{v}}{2\pi e}\sigma_{\rm tr}
\end{equation}
where $k_{\rm F}$ is Fermi wave-vector, $\vec{v}$ is the velocity of electron and $\sigma_{\rm tr}$ is the cross-section of scattering around the impurity.
The dipolar potential is given by $\delta \phi(r) =-\frac{\vec{p}\cdot \hat{x}}{r}$ where $\hat{x}$ is the unit vector pointing from the impurity along the current direction, and the direction of RRD is always opposite to the current direction.

Apart from the contribution of RRD, another effect was also proposed by Landauer~\cite{landauer1974driving,landauer1976spatial} where the influence of charged impurities on the local electron density would in turn contribute to the potential distribution and this effect was termed as CDM effect.
CDM effect also gives rise to an additional dipolar potential distribution, and the CDM dipole moment proposed by Landauer is given by~\cite{landauer1974driving,landauer1976spatial}:
\begin{equation}\label{EQ2}
  \vec{p}=\frac{3\pi}{4}\frac{Z \vec{E_0}}{3n_0+\beta \delta n}
\end{equation}
where $Z$ is the charge of impurity, $E_0$ is the macroscopic electric field, $n_0$ is the unperturbed carrier density, $\delta n$ is the carrier density perturbation introduced by the impurity and $\beta$ is the geometric factor.
It clearly shows that the CDM dipole direction is related to the sign of Z, i.e. the charge polarity of the impurity. As the electron current passing through the carrier density modulated area (Fig.~\ref{Fig1}(d) and (g)), the continuity of current requires  decreasing or increasing electron velocity (Fig.~\ref{Fig1}(e) and (h)) by local field which is generated by the self-build dipole - that is the CDM dipole~\cite{sorbello1997theory}. In contrast to RRD, the direction of CDM dipole of positively charged impurity is parallel to the current as shown in Fig.~\ref{Fig1}(f), while it is antiparallel to the current for the negatively charged impurity(Fig.~\ref{Fig1}(i)).

Experimentally demonstrating Landauer's theory and imaging those dipoles require the measurement of potential distribution at atomic scale and pose great challenges. Only after the recent progress in scanning tunneling potentiometry (STP)~\cite{muralt1986scanning,muralt1987scanning,chu1989scanning,kirtley1988direct,bannani2008local,krebs2023imaging,homoth2009electronic} and measurement of scattering effect of mobile carriers at atomic scale \cite{bannani2008local,hus2017spatially,willke2017electronic,hus2017detection} does this become possible. The experimental evidences of RRD have been revealed in bismuth thin film \cite{briner1996local,feenstra1998search}, atomic step edges of graphene \cite{wang2013local,willke2015spatial} and topological insulator thin films\cite{lupke2017electrical}. However, evidence of the current induced dipoles of atomic point-like defect is still absent. Moreover, so far, to our best knowledge, there is no direct experimental evidence of electric dipole induced by CDM effect. Here, by employing home-made low temperature multi-probe scanning tunneling microscope\cite{li2019construction}, where two contact probes build potential gradient on the sample surface and the scanning probe measures local electrochemical potential (Fig.~\ref{Fig1}(k)), we reveal two distinct electric dipoles around atomic impurities whose direction is dependent on the charge polarity of the impurity - the direct evidence of CDM dipole. Furthermore, we manipulate the directions of these electric dipoles by \textit{in situ} redirecting current. Our experiments directly visualize the potential fluctuation, i.e. current induced dipoles, around atomic point-like defect and provide unambiguous evidence for the existence of CDM dipole at atomic scale.


\section{Electric dipole of single positively charged impurity}

\begin{figure*}[htbp]
   \centering
   \includegraphics[width=0.8\textwidth]{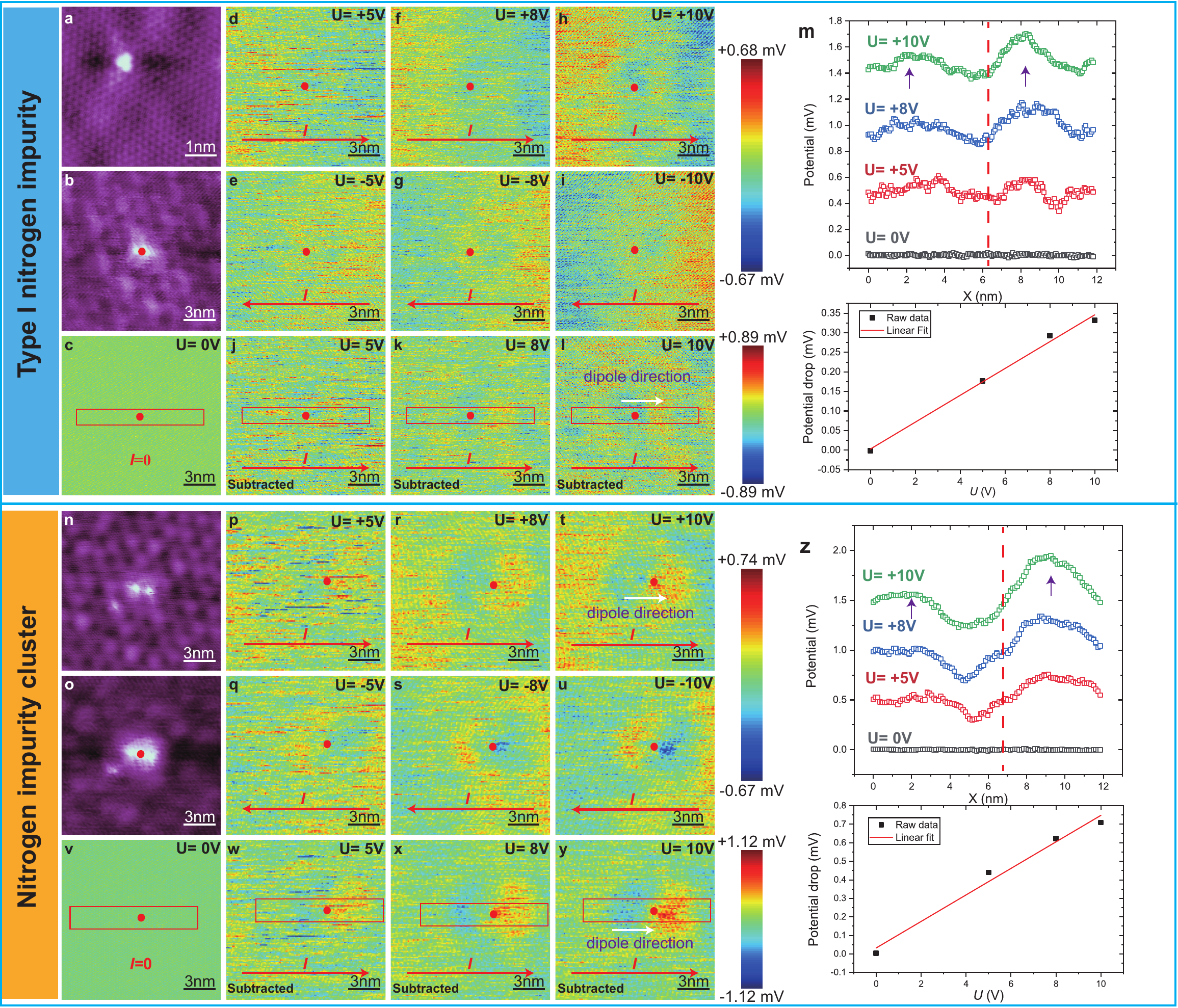}
    \caption{Current induced dipoles of positively charged impurities. a, Atomically resolved STM image of type-\uppercase\expandafter{\romannumeral1} nitrogen impurity, 5  $\rm \times$ 5 nm, $V_{\rm b}$ = 50 mV, $I_{\rm t}$ = 500 pA. b, Simultaneous topography image of STP measurement (15  $\rm \times$ 15 nm, alternating voltage: $V_{\rm{ac}} = $ 5 mV and $I_{\rm t}$ = 100 pA). c-i, Potential distribution with $U$ = 0 V, $\rm \pm$ 5 V, $\rm \pm$ 8 V and $\rm \pm$ 10 V.  j-l, Subtracted potential map denoted as $\rm P_{sub}(r, +U)-P_{sub}(r, -U)$ (see Extended Data Fig. 1 for details) with $U$ = 5 V, 8 V and 10 V.  m, Upper panel: averaged potential profile in the red box in (c, j-l). The red dashed line indicates the position of type-\uppercase\expandafter{\romannumeral1} nitrogen impurity and the distance between the purple arrows indicates the size of the electric dipole. Lower panel: potential drop across type-\uppercase\expandafter{\romannumeral1} nitrogen impurity as a function of $U$. 
n, Topography of nitrogen impurity cluster where two closely located type-\uppercase\expandafter{\romannumeral1} nitrogen impurities are resolved, 15 $\rm \times$ 15 nm, $V_{\rm b}$ = 50 mV, $I_{\rm t}$ = 500 pA. o, Simultaneous topography image of STP measurement for the nitrogen impurity cluster(15 $\rm \times$ 15 nm, alternating voltage: $V_{\rm{ac}} = $ 5 mV and $I_{\rm t}$ = 100 pA). p-v, Potential distribution with $U$ = $\rm \pm$ 5 V, $\rm \pm$ 8 V, $\rm \pm$ 10 V and 0 V.  w-y, Subtracted potential map with $U$ = 5 V, 8 V and 10 V. z, Upper panel: averaged potential profile in the red box in (v-y). The red dashed line indicates the position of nitrogen impurity cluster and the size of the electric dipole is indicated by the distance between the purple arrows. Lower panel: potential drop across nitrogen impurity cluster as a function of $U$.
The red arrows in all figures represent the direction of current and the white arrows indicate the directions of electric dipoles. The red dots represent the centers of the impurities.}\label{Fig2}
\end{figure*}

Substituted nitrogen atom, a type of well studied defect in graphene~\cite{zhao2011visualizing,joucken2019electronic,telychko2022gate,mallada2020atomic,sforzini2016structural,joucken2021sublattice}, is an ideal point-like defect to study the local current induced dipole. Due to the ability of nitrogen atoms to incorporate $\rm sp^2$ configuration and similar bond length of C-C and C-N~\cite{telychko2022gate}, nitrogen substitution keeps the honeycomb lattice of graphene unchanged.  Furthermore, nitrogen substitution in graphene lattice is equivalent to introducing a local artificial proton~\cite{telychko2022gate,mallada2020atomic}, and  modulates the local carrier density which is required to observe the CDM effect.

Nitrogen impurities are introduced by exposing atomic flat bilayer graphene films (Fig.~\ref{Fig1}(l)) to nitrogen atom beams produced by a radio frequency plasma source (see Methods).
Due to the effects of buffer layer~\cite{sforzini2016structural}, most nitrogen impurities reveal themselves as point-like substitutional impurities as is shown in the inset of Fig.~\ref{Fig1}(l).
Two type substitutional nitrogen impurities are resolved (Extended Data Fig. 2).
Although both of them are in triangular shape (Fig.~\ref{Fig2}(a)), their tunneling spectra show distinct features.
One of them exhibits two resonance peaks at around 95 meV and 165 meV (Extended Data Fig. 2 (d)) and we name this kind of nitrogen impurities as type-\uppercase\expandafter{\romannumeral1} and the other type-\uppercase\expandafter{\romannumeral2}, which is verified by first principle calculations(Extended Data Fig. 2(c)).
Unlike type-\uppercase\expandafter{\romannumeral1}, type-\uppercase\expandafter{\romannumeral2} nitrogen impurity shows a broad resonance peak which is more prominent in the large bias scale d$I$/d$V$ (Extended Data Fig. 2 (e)).
The difference between type-\uppercase\expandafter{\romannumeral1} and type-\uppercase\expandafter{\romannumeral2} impurities in d$I$/d$V$ spectra is due to two different substitutional position in bilayer graphene (A, B sublattice)~\cite{joucken2021sublattice}.
The resonance features of both type-\uppercase\expandafter{\romannumeral1} and -\uppercase\expandafter{\romannumeral2} nitrogen impurities suggest donor like behavior with attracting potential for electrons~\cite{joucken2012localized,pereira2008modeling,joucken2021sublattice}.
Since nitrogen substitution in graphene lattice serves as an attractive center, it leads to local electron doping effect~\cite{zhao2011visualizing,joucken2012localized,joucken2015charge} which is also in agreement with our data (Extended Data Fig. 2).

Then, we studied the potential distribution around both types of nitrogen impurities by STP measurement. Figure~\ref{Fig2}(b) presents the simultaneously recorded topography during potential measurement, and
Fig.~\ref{Fig2}(c) exhibits the potential map with $U$ = 0 V.
In Fig.~\ref{Fig2}(d-i) we present the potential map around type-\uppercase\expandafter{\romannumeral1} nitrogen impurity with $U$ = $\rm \pm$ 5 V, $\rm \pm$ 8 V and $\rm \pm$ 10 V.
Potential distribution around type-\uppercase\expandafter{\romannumeral1} impurity reveals a dipolar feature and the contrast is stronger with increasing $U$.
The dipolar feature is even better visualized in the subtracted potential maps as shown in Fig.~\ref{Fig2}(j-l)(see detailed process in Extended Data Fig. 1).
The subtracted maps also help to eliminate the influence of thermal voltage~\cite{willke2015spatial}.
Most importantly, the direction of the electric dipole (the white arrow in Fig.~\ref{Fig2}(l)) is aligned with local current, which indicates that this current induced dipole is dominated by the CDM effect.
The upper panel of Fig.~\ref{Fig2}(m) presents the averaged potential profile in the red box in Fig.~\ref{Fig2}(c, j-l).
The potential drop (from peak to dip) across the type-\uppercase\expandafter{\romannumeral1} nitrogen impurity, which increases from 0.151 mV ($U$ = 5 V) to 0.332 mV ($U$ = 10 V), has a linear relationship with $U$ as indicated in Fig.~\ref{Fig2}(m) lower panel. The local current density here is estimated upto 25 A/m for $U$ = 10 V, which is considerably higher than that in previous studies~\cite{ji2012atomic,willke2015spatial}. In addition, to exclude the possibility of intrinsic potential fluctuations of graphene, we conducted the same controlled experiment on a clean graphene area and no visible potential dipole has been observed(Extended Data Fig. 7).

In contrast to type-\uppercase\expandafter{\romannumeral1} nitrogen impurity, type-\uppercase\expandafter{\romannumeral2} nitrogen impurity induced dipole is relatively weaker(Extended Data Fig. 3).
The reason why type-\uppercase\expandafter{\romannumeral2} nitrogen hosts a relatively weaker dipole is due to its smaller effective charge, which is evidenced by the fact that type-\uppercase\expandafter{\romannumeral1} impurity's influence on local density of states spans about 3.5 nm while type-\uppercase\expandafter{\romannumeral2} spans only about 2 nm (Extended Data Fig. 4).



\section{Electric dipole of positively charged impurity cluster}
Next, we studied the potential distribution around impurity cluster, where the stronger scattering effect and more pronounced potential variation is expected due to the increased effective charge. The impurity cluster we have chosen in our experiment consists of two close type-\uppercase\expandafter{\romannumeral1} nitrogen impurities as shown in Fig.~\ref{Fig2}(n), whose center is marked by a red dot (Fig.~\ref{Fig2}(o)). As the current passing through, a prominent dipolar potential distribution around the cluster has been observed as presented in Fig.~\ref{Fig2}(p-u), and the signal is significantly enhanced as $U$ increases from 5 V to 10 V. After reversing current direction by changing the polarity of $U$, the dipole direction is also reversed, as shown in Fig.~\ref{Fig2}(q, s, u). The subtracted potential is even more prominent as shown in Fig.~\ref{Fig2}(w-y). The direction of the dipole is aligned with local current resembling single type-\uppercase\expandafter{\romannumeral1} nitrogen impurity. The center of the electric dipole coincides with the center of impurity cluster as shown in Fig.~\ref{Fig2}(o-y).


The average potential profiles from the red box in Fig.~\ref{Fig2}(v-y) are shown in the upper panel of Fig.~\ref{Fig2}(z), which reveals larger potential drop compared with single type-\uppercase\expandafter{\romannumeral1} nitrogen impurity.
The potential drop across the impurity cluster shows the similar linear relationship with $U$ (lower panel of Fig.~\ref{Fig2}(z)) and increases from 0.433 mV ($U$ = 5 V) to 0.710 mV ($U$ = 10 V). Compared with single type-\uppercase\expandafter{\romannumeral1} nitrogen impurity, the increased potential drop reveals a stronger CDM effect from the impurity cluster as expected.

\section{Electric dipole of negatively charged impurity}

\begin{figure*}[htbp]
   \centering
   \includegraphics[width=0.7\textwidth]{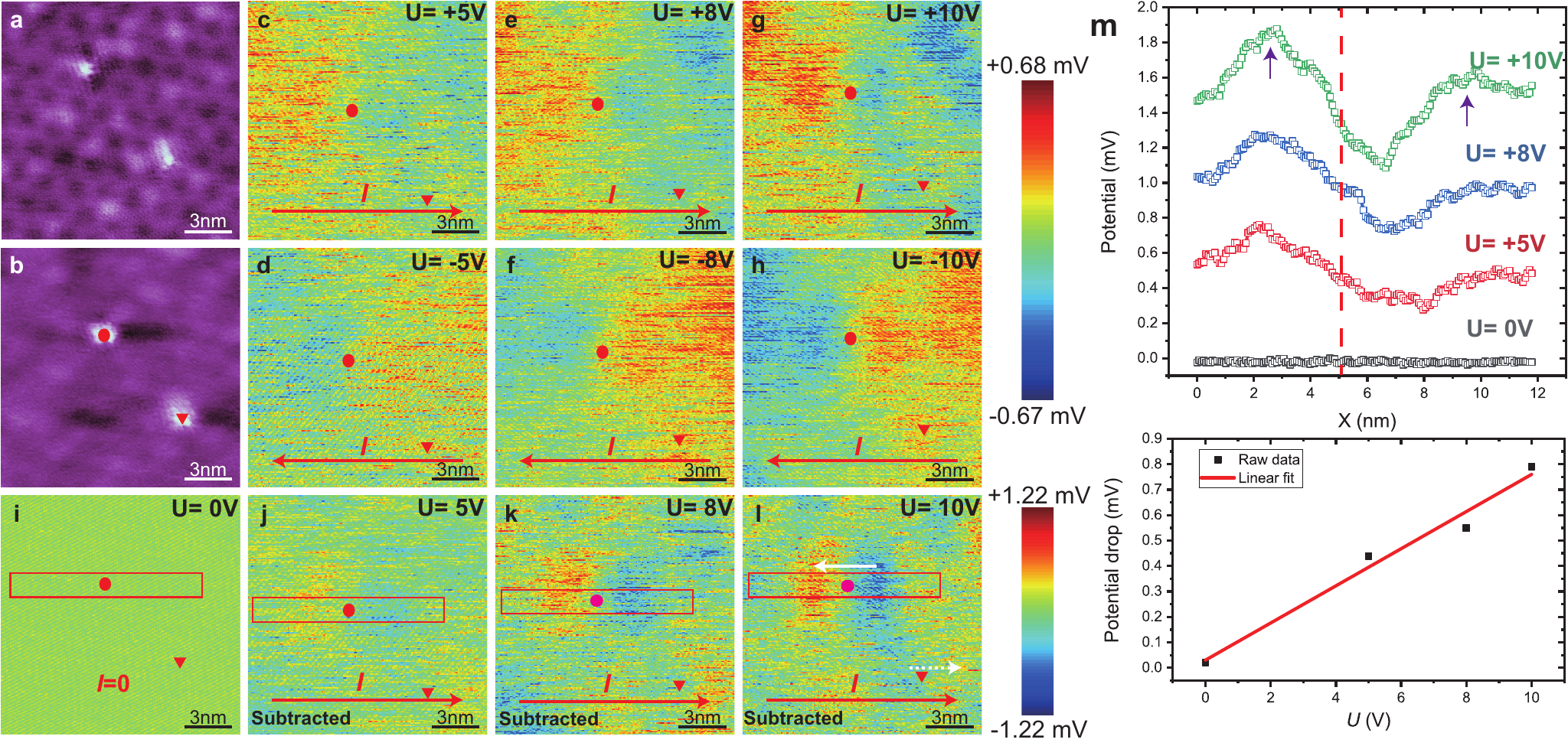}
    \caption{Current induced dipole of negatively charged impurity. a, Topography of substrate dopant at a low bias, 15 $\rm \times$ 15 nm, $V_{\rm b}$ = 50 mV, $I_{\rm t}$ = 500 pA.  b, Simultaneous topography image of STP measurement for negatively charged impurity (15 $\rm \times$ 15 nm, $V_{\rm{ac}} = $ 5 mV and $I_{\rm t}$ = 100 pA, $U$ = 0). Apart from the substrate dopant (represented by the red dot), there is also a type-\uppercase\expandafter{\romannumeral1} nitrogen impurity on the lower-right place as  indicated by the downwards triangle. c-i, Potential distribution with $U$ = $\rm \pm$ 5 V, $\rm \pm$ 8 V, $\rm \pm$ 10 V and 0 V. j-l,  Subtracted potential map with $U$ = 5 V, 8 V and 10 V.  m, Upper panel: averaged potential profile in the red box in (i-l). The red dashed line indicates the position of substrate dopant and the distance between purple arrows shows the size of the electric dipole. Lower panel: potential drop across substrate dopant as a function of $U$. 
The red arrows in all figures represent the direction of current and the white arrows indicate the direction of electric dipoles. The white dashed arrow in (l) represents the dipole direction of type-\uppercase\expandafter{\romannumeral1} nitrogen impurity.}\label{Fig3}
\end{figure*}

So far, the potential distribution around positively charged impurity, where carrier density is enlarged, has been investigated. Next, we studied the potential distribution around a negatively charged impurity, which reduces the nearby carrier density. The substrate dopant provides a perfect point-like defect with negative charge.
Substrate dopant in SiC, which is identified as a nitrogen atom substituting a carbon site, exhibits a repulsive potential for electrons in graphene and locally hole-doping effect\cite{zhang2021substrate}.
The STM topography of substrate dopant presents itself as a dark pit at a large bias (Extended Data Fig. 4(i)) and a bright protrusion at a low bias (Fig.~\ref{Fig3}(a)). Near the substrate dopant there is a type-\uppercase\expandafter{\romannumeral1} nitrogen impurity as indicated by the downward triangle in Fig.~\ref{Fig3}(b).
The d$I$/d$V$ spectra (Extended Data Fig. 4(k)) exhibit a peak around 4 V on both substrate dopant and bilayer graphene,  and the peak position in the spectrum of substrate dopant is upward shifted comparing with the bilayer graphene.
Higher resolution d$I$/d$V$ spectra around Fermi level (Extended Data Fig. 4(l)) also show that the Dirac point on substrate dopant moves upto -360 meV comparing with -410 meV of bilayer graphene. Both shifts in spectra suggest the reduction of electron density effect on the graphene by the substrate dopants, which is consistent with previous work~\cite{zhang2021substrate} (detailed spatially resolved d$I$/d$V$ in Extended Data Fig. 4(m-o)).


Then, we investigate the current induced potential distribution around substrate dopant and the results are presented in Fig.~\ref{Fig3}(c-h). For positive $U$, the negative potential gradient along the x axis can be clearly seen in Fig.~\ref{Fig3}(c, e, g), while the reversed potential gradient can be observed in Fig.~\ref{Fig3}(d, f, h) with negative $U$. In addition to that, a dipolar potential distribution superimposes on these smooth potential backgrounds. Due to the same directions of dipolar field and the backgrounds, the dipole features induced by negatively charged impurity are not easily identified as the above positively charged impurities. However, the subtracted potential maps of Fig.~\ref{Fig3}(j-l) clearly reveal potential dipoles which are stronger with increasing $U$ and the direction of the dipoles (white arrow in Fig.~\ref{Fig3}(l)) are anti-aligned with local current direction, which are reversed comparing with above positively charged impurities. The potential profile in the upper panel of Fig.~\ref{Fig3}(m) reveals an increasing potential drop from 0.439 mV to 0.790 mV as $U$ is enlarged from 5 V to 10 V, which is much more prominent than the case of single type-\uppercase\expandafter{\romannumeral1} nitrogen impurity.
For the negatively charged impurities, the directions of both RRD and CDM dipole are anti-aligned with local current, and the combination of these two effects leads to a stronger potential dipole for substrate dopant.
In stark contrast to the substrate dopant, the potential response around nearby type-\uppercase\expandafter{\romannumeral1} impurity marked by a red triangle in Fig.~\ref{Fig3}(j-l) is relatively weaker, and more importantly the dipole direction (the dashed white arrow in Fig.~\ref{Fig3}(l)) is aligned with local current.
The observation of two reversed dipoles of above both positively and negatively charged impurities further demonstrates the CDM effect, which is dependent on impurity charge polarity.


Theoretically, the size of the dipole is determined by the screening length of the system~\cite{landauer1974driving,landauer1976spatial,sorbello1997theory}. In our experiments, a similar lateral size of about 7 nm for these current induced dipoles has been revealed by the profiles of potential in Fig.~\ref{Fig2}(m), Fig.~\ref{Fig2}(z) and Fig.~\ref{Fig3}(m), which is consistent with the screening length, around 10 nm, of our epitaxial graphene~\cite{wong2017spatially}. The size of the dipole is much smaller than the estimated mean free path, about 50 nm, in our sample(see Methods). Furthermore, due to the screening effect, the oscillation\cite{sorbello1981residual,zwerger1991exact} of dipolar potential distribution with half the Fermi wavelength, which is 4.55 nm from our quasiparticle interference measurement (Extended Data Fig. 5), hasn't been observed in our experiment. More importantly, a diffusive model cannot simulate the dipolar potential observed in our experiment(Extended Data Fig. 9).

\begin{figure*}[htbp]
   \centering
   \includegraphics[width=0.65\textwidth]{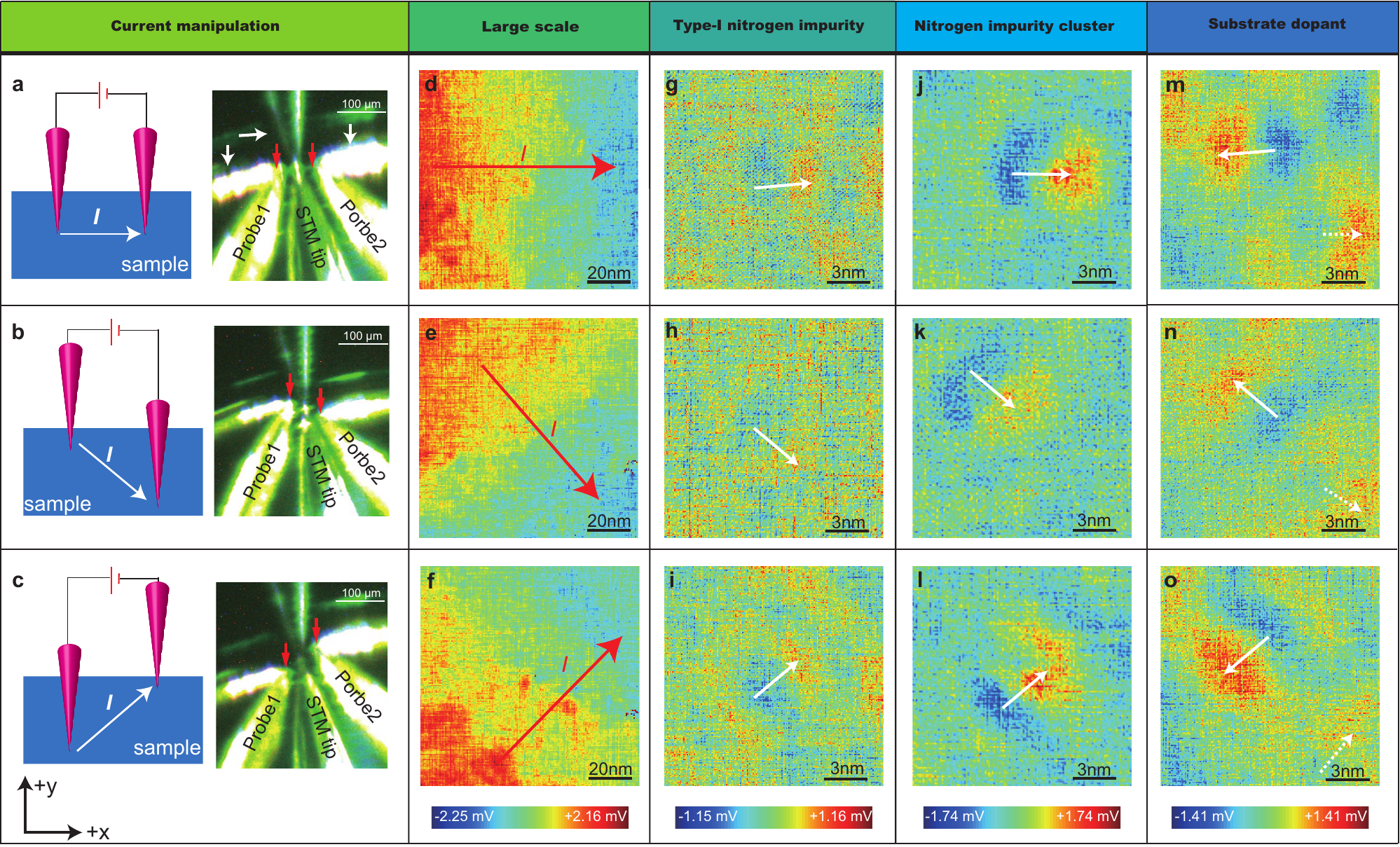}
    \caption{Manipulation of electric dipoles by local current. a-c, Left panel: schematics for tip arrangements to change the directions of local current. Right panel: optical images of real tip arrangements and the red arrows indicate the contact positions of the tips. The white arrows in the right panel of (a) mark the reflections of real probes. d-f, Reconstructed potential distribution maps ($\rm P^{x+y}(r, 10 V)$, 100 $\rm\times$ 100 nm). The red arrows indicate the directions of local current which are consistent with the schematics in (a-c). g-o, Subtracted and reconstructed potential maps of dipoles around type-\uppercase\expandafter{\romannumeral1} nitrogen impurity (g-i, see Extended Data Fig. 1 for details), nitrogen impurity cluster (j-l) and substrate dopant (m-o).}\label{Fig4}
\end{figure*}

\section{Manipulation of electric dipoles by current}
Although the dipolar potential distribution of impurities has been presented above, the entanglement of the dipole direction with the local current direction, which is one of the essential properties of current induced dipoles, hasn't been fully demonstrated.  In order to obtain various local current directions, we arrange two contact tips, between which a voltage has been applied, in three configurations as schematically shown in the left panels of Fig.~\ref{Fig4}(a-c), and the real optical images of contacted tips are presented in the right panels. In Fig.~\ref{Fig4}(a), the current is almost ideally along +x direction, while in Fig.~\ref{Fig4}(b)(\ref{Fig4}(c)), the current consists of both +x and -y (+y) direction components. To better capture the potential distribution along y axis, we scan the same region with the fast scan directions along both x and y axes and then we add them together (reconstructed potential distribution, denoted as $\rm P^{x+y}(r, U)$). The same subtraction procedure mentioned before is also used to eliminate the ohmic potential background and the thermal voltage (subtracted and reconstructed potential distribution, see Extended Data Fig. 1). In order to precisely determine the local current directions, the large scale reconstructed potential distribution maps ($\rm P^{x+y}(r, 10 V)$) in Fig.~\ref{Fig4}(d-f) have been used and the potential gradient of these maps indicate the local current directions as shown by the red arrows, which are consistent with the schematics in Fig.~\ref{Fig4}(a-c)(also see Extended Data Fig. 8). In Fig.~\ref{Fig4}(g-i), we present the dipolar potential distribution around type-\uppercase\expandafter{\romannumeral1} nitrogen impurity with three different current directions. In Fig.~\ref{Fig4}(g), the electric dipole is along +x direction aligning with local current direction as indicated in Fig.~\ref{Fig4}(a). After adding the current component along -y direction as indicated in Fig.~\ref{Fig4}(b), the electric dipole is clockwise rotated and points between +x and -y directions as shown in Fig.~\ref{Fig4}(h). Similarly, after adding +y current component as in Fig.~\ref{Fig4}(c), the electric dipole is anticlockwise rotated comparing with Fig.~\ref{Fig4}(g) and points between +x and +y directions as shown in Fig.~\ref{Fig4}(i). The rotation behavior and pointing directions of the dipole of impurity cluster along with the local current is similar to the type-\uppercase\expandafter{\romannumeral1} nitrogen impurity as in Fig.~\ref{Fig4}(j-l). The realignment of the dipole direction of substrate dopants along the current direction is also observed as shown in Fig.~\ref{Fig4}(m-o) except that dipolar direction is always anti-aligned with local current. The opposite dipolar alignments with local current of positively charged type-\uppercase\expandafter{\romannumeral1} nitrogen impurity/impurity cluster and negatively charged substrate dopants are even more clearly demonstrated in Fig.~\ref{Fig4}(m-o), where dipoles are marked by solid and dashed arrows. The rotated direction of dipolar potential distributions aligning or anti-aligning along local current not only further confirm the current induced dipoles of point-like defect and the existence of CDM effect in our results, but also provide an approach to tune the local dipolar potential distribution at atomic scale.





\section{Summary and outlook}

In conclusion, by using low-temperature multi-probe scanning tunneling microscopy, we have discovered current induced dipoles around atomic impurities in epitaxial graphene thin films. Two type dipoles have been revealed: one, induced by positively charged impurities, is aligned with local current, and the other, generated by negatively charged substrate dopants, is anti-aligned with local current. This charge polarity dependent dipolar potential distribution unambiguously demonstrates the existence of CDM effect because the RRD is always  opposite with local current and only the direction of CDM dipole depends on the charge polarity of impurities. In fact, what we have observed in our experiment is the combination of both RRD and CDM dipole. Moreover, by manipulating local current direction with contact probes, we experimentally demonstrate that the dipole orientation is redirected by local current, which is another essential property of current induced dipoles. Our work extends the capability of probing the local potential distribution down to a single atomic impurity limit and demonstrates the validity of Landauer's predictions of current induced dipoles, especially the CDM dipole, which are beyond the semiclassical theory. We expect that many quantum transport phenomena related to dissipationless topological states, Cooper pairs of superconductors and other interesting many-body states can also be investigated at atomic scale by the same technique we have demonstrated here.

\section{Methods}
\subsection{Sample preparation}
Bilayer graphene films were grown on 6H-SiC(0001) surface by thermal decomposition approach.
The 6H-SiC(0001) substrate was degassed at 600 $\rm ^{\circ}$C over night in molecular beam epitaxy (MBE) chamber (base pressure: $\rm 1\times10^{-10}$ Torr).
Next, the substrate was heated upto 850 $\rm ^{\circ}$C under Si flux, and then was flashed to 1300 - 1400  $\rm ^{\circ}$C for several times with the lasting time of 5 to 10 minutes.

Nitrogen dopants were implanted by exposing pristine graphene to nitrogen atom flux produced by a radio frequency plasma source (250 W) with sample held at 850 $\rm ^{\circ}$C for 30 minutes.
After exposing, subsequent high temperature annealing at 1300 $\rm ^{\circ}$C for 60 minutes produces a clean nitrogen doped graphene.
\subsection{STP measurement}
Our home-made multi-probe STM~\cite{li2019construction}, which is combined with a MBE system, has been used for STP~\cite{bannani2008local}.
During the STP measurement, we use two contact probes to provide current through the sample and the tip distance is monitored through optical camera.
In our experiment, the tip distance is estimated to be 50 {\textmu}m and the electric current dominantly flows through the graphene layer rather than the conducting substrate(Extended Data Fig. 10)~\cite{hasegawa2002electronic}.
The STM probe is used for both conventional STM/S characterization and STP measurement.
All STM and STP experiments are conducted at 77 K with Pt-Ir alloy tips, which are treated and characterized on Ag(111) surface.
The optical image of STP configuration is presented in Fig.~\ref{Fig1}(k) where two contact probes are used to apply voltage $U$ onto the surface of graphene.
The STP measurement records simultaneously both topography and electrochemical potential with atomic spatial resolution.

\subsection{Estimation of sheet resistance}
The sheet resistance can be estimated by $\rho_{\rm sheet}=\frac{E}{j}$, where $E$ is local electric field and $j$ is the current density.
In Extended Data Fig. 6, we present a large scale potential drop on bilayer graphene  with $U$ = 10 V.
In order to estimate the sheet resistance of bilayer graphene, we first averaged the potential drop in the blue box in Extended Data Fig. 6 (a).
The potential profile is shown in Extended Data Fig. 6 (c) and the linear fit gives an electric field of 1.10 $\times$ $\rm 10^4$ V/m.
At $U$ = 10 V, the estimated current density is given by~\cite{ji2012atomic}: $j = \frac{2 I}{\pi d}$, where $d$ is the tip distance (around 50 \textmu m in our experiment, and the contact resistance is measured to be 5000 $\rm \Omega$ ) and hence the estimated current density is around 25 A/m.
Using $\rho_{\rm sheet}=\frac{E}{j}$, the estimated sheet resistance of bilayer graphene at 77 K is $\rho_{\rm sheet}^{\rm BLG}$ = 440 $\rm \Omega/\Box$ and the results are in agreement with previous reports~\cite{jobst2010quantum,jobst2011transport,sinterhauf2020substrate}.

\subsection{Estimation of mean free path}
Using Drude model, we estimate the mean free path of our epitaxial bilayer graphene.
The conductivity is calculated by $\vec{j}=\sigma\vec{E}$ as in previous section.
The scattering time is given by $\tau=\frac{\mu E_F}{e v_F^2}$, where $\mu=\frac{\sigma}{ne}$ is the mobility and $E_F$ is the Fermi energy, $v_F$ is the Fermi velocity of Dirac electrons, $e$ is the charge of electron and $n$ is the electron density of the system.
The mean free path, which is given by $l=v_F\tau$~\cite{tan2007measurement}, is around 50 nm in agreement with previous report~\cite{weingart2009low}.

\begin{acknowledgments}
We thank S. Wu, X.-C. Ma, K. He, L.-L. Wang, C.-L. Song, W. Li, D. Zhang and X. Feng for the stimulated discussion. This work is financially supported by the Ministry of Science and Technology of China (Grants No. 2018YFA0305603, 2021YFE0107900) and the National Natural Science Foundation of China (Grants No. 11427903, 52388201).
\end{acknowledgments}



%

\end{document}